# ENSINO DE FÍSICA COM ENFOQUE CTS: CONTRIBUIÇÕES ENTRE CIÊNCIAS E A LEI 10.639/2003

**Marcelo Vilela da Silva**
Licenciado em Física-IFRJ, Doutorando em Eng. Nuclear-COPPE/UFRJ, marcelodomoba@gmail.com
**Eduardo Folco Capossoli**
Departamento de Física e Mestrado Profissional em Praticas de Educação Básica (MPPEB), Colegio Pedro II, 20.921-903 - Rio de Janeiro-RJ - Brazil, eduardo_capossoli@cp2.g12.br

**RESUMO**
O presente trabalho se propõe a desenvolver ações formativas que contemplem as relações compreendidas entre ciência, tecnologia e sociedade (CTS), no âmbito da disciplina de física - nos ensinos fundamental e médio, com intuito de trabalhar transversalmente a Lei 10.639/2003, e, portanto, os conteúdos de história e cultura africana e afro-brasileira, nos currículos escolares. Além disso, o artigo tem como objetivo a superação dos desafios encontrados e enfrentados no interior do sistema educacional a fim de tornar possível, aos estudantes do ensino médio, um conhecimento a respeito das contribuições que as sociedades africanas deram para a ciência e a tecnologia. O procedimento metodológico incluiu um questionário escrito pré e pós-teste, ambos, com os estudantes. Os resultados encontrados foram analisados qualitativamente, e demonstraram considerável relevância para se trabalhar, em sala de aula, as relações étnico-raciais.

**Palavras-chave:** Ensino de Física; Ciências-Tecnologia-Sociedade (CTS); Lei 10639/2003.

# TEACHING PHYSICS WITHIN STS APPROACH: CONTRIBUTIONS BETWEEN SCIENCES AND ACT 10.639/2003

**ABSTRACT**
This paper is aimed to develop formative actions about Science Technology and Society (STS) relations in the teaching of Physics in order to integrate Act 10.639/2003, which includes African, and Afro-Brazilian history and culture in school syllabuses. Besides we also aim to overcome the challenges found in the educational system and to enable students to learn about the contributions made by African societies regarding to the Science and technology. The methodological procedure included a written questionnaire, pre- and post-test, both with the students. The results have been proven relevant for the teaching of racial relations in the classroom.

# 1. INTRODUÇÃO

Esta pesquisa é um recorte da Produção Acadêmico Final (PAF) apresentada ao programa de Residência Docente (PRD), vinculado à Pró-Reitoria de Pós-Graduação, Pesquisa, Extensão e Cultura (PROPGPEC), do Colégio Pedro II, como requisito para obtenção do título de Especialista em Docência da Educação Básica (SILVA, 2018).

Além do PAF, foi elaborado um Produto Educacional (PE), apresentando aos estudantes do Ensino Médio (EM) - um guia envolvendo conhecimentos da física, da metrologia e da radiação, a saber: massa, peso, unidades de medidas, energia nuclear, radiações ionizantes e não ionizante, atrelados ao conhecimento advindo da história e ciência africana, com objetivo de desmistificar concepções errôneas de que estes povos seriam primitivos.

O PE foi aplicado em uma escola da rede estadual de ensino, localizada no município do Rio de Janeiro, entre os anos de 2015 e 2016. Adicionalmente, foi elaborado um questionário escrito (diagnóstico) sobre assuntos de física nuclear, com enfoque na proteção radiológica, envolvendo, particularmente, alguns assuntos relacionados ao conhecimento de africanidades. Nesse sentido, o material em questão buscou auxiliar estudantes e professores, proporcionando a eles uma ferramenta didático-metodológica que contribuísse no processo de ensino-aprendizagem da história e da ciência, tendo como foco a aplicação da lei 10.639/03, contemplando aspectos da Física Moderna e Contemporânea (FMC).

O Comitê Científico das Nações Unidas para os Efeitos da Radiação Atômica (UNSCEAR), as aplicações das radiações ionizantes são utilizadas em todo o mundo como instrumentos indispensáveis para resguardar e melhorar a saúde humana, porém representam as maiores fontes de exposição à radiação (UNSCEAR, 2000; 2008). Além disso, o uso da radiação ionizante e não ionizante, tem sido amplamente empregado no âmbito da medicina, em procedimentos intervencionistas (IAEA, 2012).

Entendemos ser absolutamente necessário aproximar os estudantes do EM de um assunto tão importante, considerando que, de certa forma, a temática está inserida no cotidiano de suas vidas, na medida em que alguns desses estudantes, eventualmente, tenham convivido e/ou conhecido pessoas que precisaram passar por procedimentos envolvendo o uso de radiação ionizante.

O currículo mínimo, da disciplina de física, do estado do Rio de Janeiro (RIO DE JANEIRO, 2012), no tópico que versa acerca das radiações ionizantes, propõe seminários para abarcar os conceitos de Física Moderna e Contemporânea (FMC). Tais iniciativas refletem a necessidade de se diversificar os modos de aprender e ensinar, buscando despertar, nos jovens, atitudes e valores que os permitam participar ativamente do mundo, de maneira autônoma e responsável, ao desenvolver de forma crítica e reflexiva suas intelectualidades.

## 2. OBJETIVOS

Este trabalho tem como objetivo principal introduzir conceitos básicos da disciplina de física, no ensino médio escolar, com atenção para seus aspectos relevantes, sem deixar de lado o enfoque em Ciência-Tecnologia-Sociedade - CTS. Para tanto, discutiremos questões como, por exemplo: o que é ciência sob diferentes aspectos; as diversas formas de se definir e pensar tecnologia; as principais formas de se interpretar a sociedade.

Para realização desse trabalho foi necessário estabelecer os seguintes objetivos específicos:

1. Deslocar para um plano social e político as questões acerca do desenvolvimento, aplicação e implicações das tecnologias e dos conhecimentos científico, com especial atenção para sociedade africana (Império Egípcio), tendo com maior destaque os conceitos de instrumentos de medida, radiações ionizantes e não ionizante;
2. Identificar experiências pedagógicas de combate ao racismo e as discriminações em sala de aula, desfazendo os equívocos sedimentados, no decorrer da história, pelos livros didáticos em torno da população negra, de sua cultura científica e tecnológica;
3. Propor novas temáticas e didáticas, no que diz respeito ao ensino da física, para se trabalhar a lei 10.639/03, encarando essa prática como ação de fundamental importância para reforçar a positivação da identidade racial negra, bem como a valorização de sua cultura.

## 3. PRESSUPOSTOS TEÓRICOS

### 3.1 Ciência, Tecnologia e Sociedade

O movimento Ciência, Tecnologia e Sociedade (CTS), emerge historicamente nas décadas de 1960 e 1970, em alguns países, tais como: EUA, Canadá, Inglaterra, entre outros, opondo-se à euforia diante dos resultados relacionados ao avanço científico e tecnológico (AULER, 1998). Recentemente, a educação tem se tornado um

dos principais campos de investigação em CTS. Conforme destacado por Pinheiro et al. (2007), devemos incentivar os estudantes da educação básica a desenvolverem seu lado reflexivo em favor de um contexto social mais justo.

De acordo com Medina e Sanmartín (1990), quando se pretende incluir o enfoque CTS no contexto educacional, é importante que alguns objetivos sejam contemplados, a saber:

- Questionar as formas herdadas de estudar e atuar sobre a natureza, as quais devem ser constantemente refletidas;
- Debater a distinção convencional entre conhecimento teórico e conhecimento prático;
- Combater a segmentação do conhecimento em todos os níveis da educação;
- Promover uma autêntica democratização do conhecimento cientifico e tecnológico, de modo que este se integre de maneira crítica às atividades das comunidades.

O desenvolvimento da ciência e da tecnologia tem possibilitado diversas transformações na sociedade contemporânea, refletindo de forma significativa nos níveis econômicos, político e, até mesmo, social, conforme apontado por Pinheiro et al. (2007). Desse modo, pode-se afirmar que a proposta de um ensino em CTS defende uma educação científica e tecnológica mais integrada ao conhecimento social, não fragmentada e tão pouco antagônica frente a outras áreas do conhecimento. Assim, acredita-se que essa perspectiva educacional possibilitará aos jovens cidadãos, o contato com habilidades e valores que os permitam tomar decisões responsáveis sobre demandas de ciência e tecnologia, na sociedade, representando, então, elementos pertinentes ao currículo, pois conforme destacam Santos e Mortimer (2002, p. 6):

> O processo de investigação científica e tecnológica propiciaria a participação ativa dos alunos na obtenção de informações, solução de problemas e tomada de decisão. A interação entre ciência, tecnologia e sociedade propiciaria o desenvolvimento de valores e ideias por meio de estudos de temas locais, políticas públicas e temas globais.

Esta estrutura conceitual, na modalidade CTS, é uma aquisição de conhecimento que enfatiza aspectos relevantes no âmbito educacional, tendo em vista a complexidade de temas relacionados aos avanços científicos.

Nesse sentido, a relação ensino-aprendizado passará a ser entendida como a possibilidade de despertar, no estudante, a curiosidade e o espírito investigador, reflexivo/crítico e também transformador da realidade em que está inserido. Com o enfoque CTS o trabalho em sala de aula passa a ter outra conotação. Professores e estudantes passam a descobrir a pesquisar, simultaneamente, a construir e/ou produzir conhecimento científico, que deixando de ser considerado unilateral, possibilita um processo de aprendizagem dialógico.

No entanto, não basta apenas julgar como importantes os pressupostos do enfoque CTS e sua consequente relevância para o contexto educacional, pois, segundo Santos e Mortimer (2002), antes de ter em mente a introdução do enfoque CTS em sala de aula, em particular no Ensino Médio, é imprescindível adotar uma análise adequada dos currículos. Com isso, o que se pretende, portanto, é promover condições para o desenvolvimento de habilidades e competências, através de estratégias pedagógicas muito bem estruturadas e organizadas. As propostas para a formação dos cidadãos precisam levar em conta os conhecimentos prévios dos estudantes, o que pode ser feito mediante a contextualização dos temas científicos (PINHEIROS et al., 2007).

## 3.2 Ensino de Física

Os Parâmetros Curriculares Nacionais (PCN) são orientações formuladas pelo governo federal, que através da articulação de diversas diretrizes, buscam estabelecer direcionamentos no processo da educação brasileira. No caso da disciplina em questão, Física, esta apresenta-se como uma espécie de conhecimento que permite elaborar modelos, investigar os mistérios do mundo 'invisível' das partículas que compõem a matéria, ao mesmo tempo em que procura desenvolver novas fontes de energia e criar diferentes materiais, produtos e tecnologias junto aquilo que defende o PCN para o ensino de Física:

> Incorporado à cultura e integrado como instrumento tecnológico, esse conhecimento tornou-se indispensável à formação da cidadania contemporânea. Espera-se que o ensino de Física, na escola média, contribua para a formação de uma cultura científica efetiva, que permita ao indivíduo a interpretação dos fatos, fenômenos e processos naturais, situando e dimensionando a interação do ser humano com a natureza como parte da própria natureza em transformação (BRASIL, 2000, p.22).

Conforme destaca a citação acima, os avanços tecnológicos são instrumentos que possibilitam uma ampliação de conhecimentos, contudo, esse conhecimento tecnológico deve ser ensinado de forma crítica e reflexiva.

A abordagem de temas sobre Física Moderna e Contemporânea (FMC), na Educação Básica, é defendida em documentos da legislação educacional brasileira

PCN+ (BRASIL, 2002) e em trabalhos de ensino de Física (OSTERMANN; MOREIRA, 2000).

Terrazan (1992) propõe, aos professores da rede escolar, alguns elementos para se pensar a problemática apontada acima, sugerindo ainda a necessidade da busca de estratégias para a inclusão de aspecto relativa ao estudo da física que desenvolveram no século passado.

Na Lei de Diretrizes e Bases da Educação (BRASIL, 1996) existem demandas educacionais direcionadas à preparação dos estudantes cujo propósito é formar sujeitos para o exercício da cidadania. Ou seja, apresentar algumas críticas ao ensino tradicional, que ainda predomina nos espaços escolares busca-se um maior envolvimento da realidade dos estudantes com o ensino da Física.

Nesse sentido, devem-se abordar os conteúdos de Física, nas escolas, como instrumento favorável à compreensão do mundo, ou seja, apresentá-los ao jovem a fim de que seja não apenas mais uma informação aleatória, deslocada da realidade, mas que a partir da tomada deste conhecimento, o estudante tenha condições de usá-lo de forma produtiva em relação a si e sua existência no mundo. Vale dizer que esse tipo de abordagem evidencia a importância da transposição didática e subsidia o processo de ensino- aprendizagem, colaborando com as formas de se pensar e agir em sala de aula.

Há teóricos (OSTERMANN; MOREIRA, 2000) que defendem alguns assuntos da Física Moderna e Contemporânea como indispensáveis à compreensão dos estudantes em relação a conhecimentos físicos pertinentes ao entendimento das novas tecnologias, visto que:

> (...) o estudo da Física deve ser compreendido pelo aluno como um processo de construção humana, inserida num contexto histórico e social, abrangendo um corpo teórico de conhecimentos científicos e tecnológicos que têm contribuído para o desenvolvimento de pesquisas que podem melhorar a qualidade de vida da sociedade (OLIVEIRA, 2006, p.21).

As discussões que vêm sendo trazidas, até aqui, neste trabalho, sinalizam para relevância de debater os conceitos e avanços da tecnologia, no contexto dos jovens, de forma a inserir ferramentas de ensino interdisciplinares necessárias ao conhecimento tecnológico e social do espaço escolar. Adiante, faremos exemplificações acerca das habilidades e competências pertinentes à presente proposta de ensino para sala de aula.

É certo afirmar que a respeito das habilidades e das competências no ensino da Física se faz necessário imprimir significados às suas estruturas, bem como aos seus conhecimentos adquiridos, de modo que "[...] os critérios que orientam a ação pedagógica deixam, portanto, de tomar como referência primeira 'o que ensinar de Física', passando a centrar-se sobre o 'para que ensinar Física'[...]" (BRASIL, 2002, p.61), visando uma formação científica e social, consequentemente, mais humanista.

O PCN+ nos apresenta alguns exemplos de habilidades e competências que se enquadram nesta proposta de trabalho, aqui, mobilizada. Estas competências em Física são: Relações entre conhecimentos disciplinares, interdisciplinares e interáreas; Ciência e tecnologia na história; Ciência e tecnologia na atualidade; Ciência e tecnologia, ética e cidadania (BRASIL (2002, p. 66-68).

De acordo com BRASIL (2002, p. 66-68), a articulação, a integração e a sistematização dos fenômenos e teorias dentro de uma ciência, possibilitam conhecer e avaliar o desenvolvimento tecnológico contemporâneo, suas relações com as ciências, seu papel na vida humana, sua presença no mundo cotidiano e seus impactos na vida social. Desta forma, estabelece um contato mais crítico e sem dogmatismo acerca dos avanços tecnológicos, de forma integrada e contextualizada, tendo como exemplo o uso de radiações ionizantes, em termos de benefícios para a saúde humana e riscos, devido às exposições em procedimentos diagnósticos.

Com relação aos temas estruturadores do ensino de Física, presentes nos PCN+ (BRASIL, 2002), seis deles foram privilegiados e delineados para integrar os objetivos a serem atingidos pela escolarização em nível médio, interessando-nos, em particular, o tema Matéria e Radiação.

Como justificativa para a escolha deste tema, ressaltamos a importância do estudo das radiações e suas interações com a matéria, tomando como base: os modelos que constituem a matéria, os tipos de radiações presentes no cotidiano, as interações das radiações com meios materiais, os efeitos biológicos das radiações, assim como medidas de proteção e radiações ionizantes. Com isso, buscamos destacar uma abordagem e uma compreensão dos fenômenos associados a essas interações e, assim, ampliar o entendimento do universo físico e microscópico.

> O cotidiano contemporâneo depende, cada vez mais intensamente, de tecnologias baseadas na utilização de radiações e nos avanços na área da microtecnologia. Introduzir esses assuntos no ensino médio significa promover nos jovens competências para, por exemplo, ter condições de avaliar ricos e benefícios que decorrem da utilização de diferentes radiações,

> compreender os recursos de diagnóstico médico (radiografias, tomografias etc.), acompanhar a discussão sobre problemas relacionados à utilização de energia nuclear [...]. (BRASIL, 2002, p.77).

De acordo com os PCN+ (2002), o amplo conhecimento acumulado na Física, ao longo da história da humanidade, não tem como ser contemplado, em sua totalidade, durante o ensino médio. Na maioria das vezes, os critérios para definir os conteúdos são restritos à própria estrutura da Física, e, portanto, não compreende o sentido mais amplo da formação cidadã desejada. Assim, conhecimentos sobre Física Médica são ferramentas utilizadas para a compreensão do mundo, de forma que a Física deve ser apresentada aos estudantes como instrumento capaz de fazer com que pense e compreenda os fenômenos naturais e cientifico de forma a estabelecer com suas principais aplicações na sociedade.

A ação norteadora pedagógica precisa deixar de ter como referência primeira "o que ensinar de Física", e passar centrar-se em "para que ensinar Física" (BRASIL, 2002, p 58). Por exemplo, ao ensinar conteúdos de eletrostática como tópicos indispensáveis à compreensão da eletrodinâmica e do eletromagnetismo, corre-se o risco de apresentá-los distantes da realidade dos estudantes. Todavia, se a referência é "para que ensinar Física", o jovem precisará lidar com situações cotidianas e reais (BRASIL, 2002).

O conhecimento adquirido sobre eletrostática respalda os conceitos básicos para compreender o magnetismo. Um exemplo disso, seria entender, como funciona um solenoide em sua concepção como agrupamento de espiras de corrente. Se partimos da premissa "para que ensinar Física", podemos transcender os conteúdos ensinados em sala de aula a fim de que os conceitos de magnetismo, possam contribuir com os conhecimentos de Física Médica, possibilitando, por conseguinte, entendimentos sobre quando pacientes são submetidos a um exame de ressonância magnética, visto que, o cilindro que o envolve é um grande solenoide que gera um campo magnético.

Portanto, há carência de propostas de ensino-aprendizagem que abordem a necessidade do jovem em adquirir competências para trabalhar com situações já vivenciadas por eles. Mas para que essa experiência se dê do modo mais democrático, como deve ser, o conhecimento a ser aprendido precisa estar comprometido com uma visão humanista abrangente, atenta ao perfil cidadão que se deseja formar. As competências enquanto eixo organizador do trabalho docente contemplam os objetivos da educação, sendo uma ferramenta para a ação dos professores (DIAS, 2013).

Algumas competências são comuns a todas as etapas do aprendizado e, outras, são específicas às fases mais avançadas - como é o caso do nível médio. Elas ganham um sentido bem determinado, por exemplo, com a introdução de modelos explicativos que constroem as abstrações indispensáveis, cabendo ao educador listar, selecionar e organizar os objetivos em torno dos quais, para ele, tem mais relevância se aplicados nas escolas (DIAS, 2013).

Para conseguir um trabalho mais integrado, as competências já foram organizadas em diferentes categorias que, no entanto, se correlacionam nos PCN+ (BRASIL, 2002), a saber: "investigações e compreensão", "linguagem física e sua comunicação" e "contextualização histórica e social".

### 3.3 A Lei 10639/2003

Esta seção tem como finalidade chamar a atenção do leitor para o recorte legal sobre o qual a pesquisa foi baseada, isto é, a inclusão da Lei 10.639 de 9 de janeiro de 2003, que altera os Artigos 26, 26 A e 79 B da Lei de Diretrizes e Bases da Educação Nacional para incluir, no currículo oficial das redes públicas e privadas de ensino, a obrigatoriedade do estudo da História e da Cultura da África e dos Africanos.

Essa regulamentação vem acompanhada do Parecer e da Resolução que instituíram as "Diretrizes curriculares nacionais para a educação das relações étnico-raciais e para o ensino de história e cultura afro-brasileira e africana", aprovados pelo Conselho Nacional de Educação (CNE), em março de 2004, e homologados pelo Ministério da Educação (MEC) em junho do mesmo ano. A resolução surge com base no Parecer CNE/CP 3/2004, e visa atender à Lei nº 10.639/2003 (BRASIL, 2003). Em 2008, a lei foi modificada, sendo acrescida a obrigatoriedade da história indígena no Brasil (lei 11.645/08).

A história da África, presente no imaginário do Brasil, se expressa, é mantida e transformada em manifestações histórico-culturais, diretamente vinculadas à uma visão de mundo "atrasada". Essas noções errôneas transitam pelo imaginário e são reforçadas constantemente pelos livros didáticos. Todavia, para pensar e ensinar cultura de matriz africana e afro-brasileira, é preciso compreender sua continuidade junto aos conhecimentos científicos e tecnológicos das significações que começaram a ser elaborados já no continente africano.

Segundo Munanga (2005), o desconhecimento da história da África, da cultura do negro no Brasil e da própria história do negro de um modo geral, por parte dos professores, é tido como um sério problema na tentativa de se implementar a lei em sala de aula, uma vez que muitos são cooptados pela ideia que se solidificou acerca mito da democracia étnico-racial. Neste sentido, o autor afirma que a:

> [...] educação é capaz de oferecer tanto aos jovens como aos adultos a possibilidade de questionar e desconstruir os mitos de superioridade e inferioridade entre grupos humanos que foram introjetados neles pela cultura "racista" na qual foram socializados (MUNANGA, 2005, p. 17).

Dessa maneira, apontar os elementos presentes nessas interpretações é também traçar estratégias para a educação das relações étnico-raciais, que tem por alvo a formação de cidadãos, mulheres e homens, empenhados em promover condições de igualdade no exercício de seus direitos sociais, políticos, econômicos, além dos direitos de ser, viver e pensar próprios, no que diz respeito aos diferentes pertencimentos étnico-raciais e sociais. Em outras palavras, persegue o objetivo precípuo de desencadear aprendizagens e ensinos em que se efetive a participação no espaço público cuja finalidade é de contribuir para uma reflexão crítica quanto às contribuições dos povos africanos.

A partir destes documentos e das considerações levantadas, enfatizamos que o ensino da física, na escola de educação básica, deve abarcar diversos saberes, não se limitando aos conhecimentos eurocêntricos. Os futuros professores que atuarão nas escolas precisam ter instrumentos que oportunizem em suas aulas - teóricas e práticas - aprendizagens que envolvam, também, saberes históricos e sociais.

Os trabalhos a seguir, mostram-se enquanto exemplos que permitem inserir conteúdos de física a partir de um currículo engajado junto à práticas efetivas que contribuem para uma educação antirracista. Esses trabalhos trazem outras narrativas e colaboram para a efetivação da Lei 10.639/2003, na medida em que colocam em destaque uma história, por vezes, silenciada ou negada na formação inicial de professores.

Tufaile (2013) discute a importância e as possibilidades da inserção de atividades práticas no ensino de Ciências, em particular na física, apresentando a "Física do Faraó", destacando sistemas de padrões e medidas usados no Egito antigo. O estudo comprova uma variedade de medidas de comprimento importante na Física, a saber: o cúbito (TUFAILE, 2013).

Noguera (2010) propõe, na área de Ciências da Natureza, Matemática e suas Tecnologias (biologia, física, química e matemática), algumas estratégias de incorporação de elementos afrocêntricos às práticas pedagógicas a serem adotadas em cada disciplina. A matéria de física, possibilita o estudo de conceitos e fenômenos acústicos, através da construção de instrumentos musicais africanos, como segue: *Kalimba*, *djembê*, e o *dundu*.

A partir do paradigma afrocentrado é relevante destacar trabalhos produzidos por africanos, conforme pesquisado por Morais (2017), em *Histórias aprisionadas de personagens como Lewis Latimer*, em que apresenta, para estudantes, uma imagem representativa do negro em momentos históricos como protagonista da sociedade. Coloca o nome de Latimer no hall dos grandes inventores e engenheiros, ao lado de personagens da ciência como o próprio Thomas Edison.

Este trabalho reforça a importância de apresentar os fundamentos apoiados na história dos povos africanos, numa linha filosófica africana, ancorados em investigações sociológicas que analisem as contribuições científicas e tecnológicas frente as diferentes sociedades africanas e afrodiaspóricas.

## 4. METODOLOGIA

Esta pesquisa trata-se de estudo exploratório de caráter qualitativo (GIL, 2010). Para conhecimento de todo processo que a envolve, é necessário dizer que à época de sua realização, o PRD, não exigia qualquer documento ou submissão de projetos à plataforma Brasil, e nem mesmo a passagem por um comitê de ética. A caracterização do campo de estudo e forma de ingresso se dá em uma escola da rede estadual, do município do Rio de Janeiro, local onde trabalha um dos pesquisadores, mais especificamente, em uma turma de 3º série do ensino médio, com 40 tempos de cinquenta minutos de hora/aula-com um total de 20 encontros, e uma turma completa de 38 estudantes, cuja faixa etária varia entre 18 e 30 anos.

Neste trabalho, a descrição da metodologia teve como proposta de atividades o levantamento de um questionário inicial (escrito) e seminários com temas variados e contextualizados, apresentados abaixo, respectivamente:

**Quadro 1 –** Questionário inicial (escrito)

> 1- Exemplifique a função do Físico/a.
> 2- Dê exemplos de física da radiação ionizante e não ionizantes.
> 3- Já ouviu falar das Pirâmides do Egito?
> 4- Quem inventou o primeiro reator nuclear?
> 5- Você conhece físicos negros/as?
> 6- Você conhece cientistas mulheres?
> 7- Você conhece cientistas africanos?
> 8- Você já ouviu falar de Aceleradores de Partículas?

Fonte: (Os autores, 2017).

**Quadro 2:** Temas dos seminários

> 1- Cheikh Anta Diop ( carbono 14).
> 2- Dê exemplos de física da radiação ionizante e não ionizantes.
> 3- As Pirâmides do Egito.
> 4- Quem inventou o primeiro reator nuclear?
> 5- Mulheres na Ciência
> 8- Aceleradores de Partículas.

Fonte: (Os autores, 2017).

**Quadro 3**: Estruturação das atividades

| ATIVIDADES 1 | |
|---|---|
| **1ª etapa:** Questionário Inicial (inscrito) | 1. O professor entrega o questionário inicial (inscrito) com perguntas que relacionam o tema abordado à experiência e ao conhecimento pessoal de cada estudante. <br><br> 2. Além de fazê-los começar a pensar sobre os assuntos da Física da Radiação, o questionário inicial servirá como um pré-teste (ou avaliação diagnóstica inicial). |
| **2ª etapa:** Abertura de aula com dois vídeos <br><br> **Vídeo 1**: "Estrelas Além do Tempo" (apresentado o trailer no seminário 1) <br><br> **Vídeo 2**: "Cientistas e inventores negros " Disponível em: https://www.youtube.com/watch?v=h0hv9fg88n8. Acesso em: 23 Set 2016. | 3. O professor faz uma introdução geral da origem da radiação e suas aplicações em tela, apresentando e discutindo dois vídeos instrucionais: o 1° abordando a temática da invisibilidade de mulheres negras na ciência; o 2° mostrando alguns cientistas e inventores negros. |
| **ATIVIDADE 2** | |
| **3ª etapa:** Discussão sobre o Questionário Inicial e um Vídeo 3 instrucional <br><br> **Video 3**: Origem do Raios X e sua aplicação. Disponível em: http://portaldoprofessor.mec.gov.br/buscaGeral.html?q=raios%20X. Acesso em:07 Jun 2006. | 4. O professor comenta as respostas apresentadas no questionário inicial, e debate os conceitos importantes construindo, com os estudantes, correlações junto aos aspectos destacados nos dois Vídeos Instrucionais. Desta forma, o professor seguirá os seguintes eixos: <br><br> ➢ O quanto as radiações ionizantes o afetaram/afetam? <br> ➢ O que pensam sobre seus efeitos biológicos? <br> ➢ O que pensam sobre os conhecimentos novos em tela? <br> ➢ Quais as relações entre os conhecimentos físicos e o problema quando pacientes são submetidos a exames Médicos para diagnóstico? <br> ➢ Energia Nuclear e natureza ondulatória. |
| | |
| **4ª etapa:** Sistematização da discussão sobre o Questionário Inicial e os Vídeos 3 e 4 Instrucional<br><br>**Vídeo 3**: Exame de ultrassografia. Disponível em: http://www.youtube.com/watch?. Acesso em 07 Agostos 16.<br><br>**Vídeo 4**: Proteção e controle de exame odontológico. Disponível em: http://www.youtube.com/watch?v=I-87BtMrjbI). Acesso em 07 jul 16. | . 5. O professor sistematiza, no quadro da sala de aula, os principais aspectos levantados pelos estudantes durante a discussão a respeito do Questionário Inicial e de um Vídeo Instrucional com abordagem em exame de ultrassonografia e, outro, com destaque à proteção e controle de exame odontológico. |
| **ATIVIDADE 3** ||
| **5ª etapa:** Texto de Apoio Didático | . 6. O professor faz uma explanação sobre aspectos centrais da Física Nuclear (Aplicações), entregando, em seguida, o Texto de Apoio Didático (SILVA, 2018).<br><br>7. Após sua leitura, cada dupla de estudantes recebe o questionário de apoio ao texto com perguntas relacionadas, diretamente, ao texto de apoio didático. |
| **6ª etapa:** Questionário de Apoio ao texto | 8. Enquanto os estudantes estiverem respondendo o questionário de apoio ao texto, o professor se movimenta pela sala de aula para esclarecer eventuais dúvidas e instigar reflexões críticas acerca do texto. |
| **ATIVIDADE 4** ||
| **7ª etapa:** Discussão sobre o Texto de Apoio Didático e seu Questionário de Apoio ao Texto | 9. O professor apresenta os seminários (1, 2 e 3), para aprofundamento do assunto, e disponibiliza temas para a futura apresentação de seminários por parte dos estudantes (SILVA, 2018). |
| **8º etapa:** Sistematização da discussão sobre o Texto de Apoio Didático e seu Questionário de Apoio ao Texto | 10. Os estudantes apresentam seminários com temas disponibilizados antecipadamente e, nesse momento, devem relacionar alguma questão às temáticas étnico- raciais (SILVA, 2018).. |

| ATIVIDADE 5 | |
|---|---|
| **9º etapa:** Considerações finais sobre a atividade diagnóstica | 11. O professor espera que os estudantes forneçam respostas positivas e façam considerações relevantes sobre as radiações e seus efeitos. Realiza duas perguntas sobre os vídeos, seguindo a sequência abaixo:<br><br>➢ Qual a importância de abordar os conceitos de radiação na escola? Justifique.<br>➢ Como podemos verificar a resposta e respectiva justificativa para a pergunta anterior?<br>➢ Proposta de pesquisa sobre temática étnico-racial. |
| **10ª etapa**: Questionário Final (escrito) | 12. O professor entrega o questionário final (igual ao questionário inicial) para que os estudantes respondam, individualmente, novamente. Dessa forma, o professor poderá avaliar se houve avanço na compreensão dos estudantes a respeito dos novos conhecimentos adquiridos, ou seja, se alguns dos objetivos almejados foram satisfatórios para cada estudante. Esse questionário final representa, então, um pós-teste (ou avaliação final). |

Fonte: (Adaptado de SANTOS, 2012, p.70-73).

## 5. APRESENTAÇÃO E ANÁLISE DOS DADOS

Nesta seção serão apresentados os resultados coletados nas atividades mencionadas na seção anterior.

Nas figuras 1, 2 e 3 verifica-se o levantamento diagnóstico, com o objetivo de aprender a respeito dos conhecimentos adquiridos pelos estudantes no que tange aos temas relacionados aos saberes científicos e tecnológicos africanos. Assim, constata-se que a relação de conhecimento sobre ciência é mínima, mas com relação ao trabalho, é alta. Podendo-se conjecturar a ideia de imagens e concepções, estereotipadas, junto à população negra. Já na figura 3, os estudantes não responderam e/ou desconhecem qualquer referência de físicos/as africanos/as.

Figura 1: Pré – teste-conhecimentos sobre a África seguindo o quadro 1.

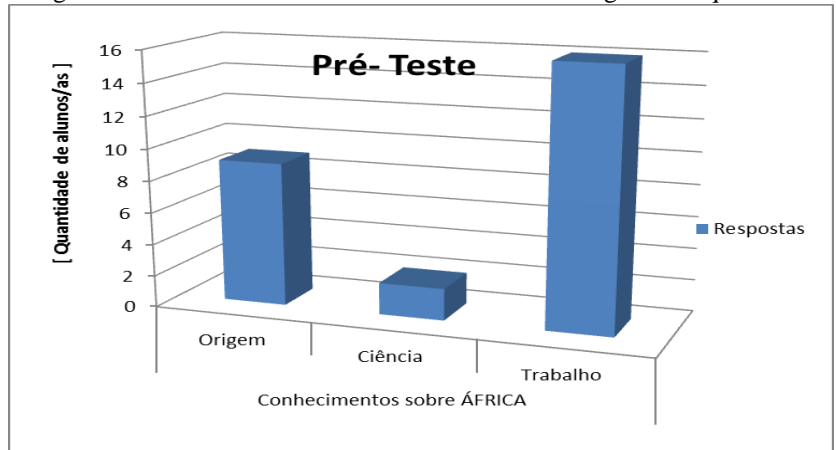

Fonte: (Os autores, 2017).

Figura 2: Pre-teste-respostas dos estudantes com o tema seguindo o quadro 1.

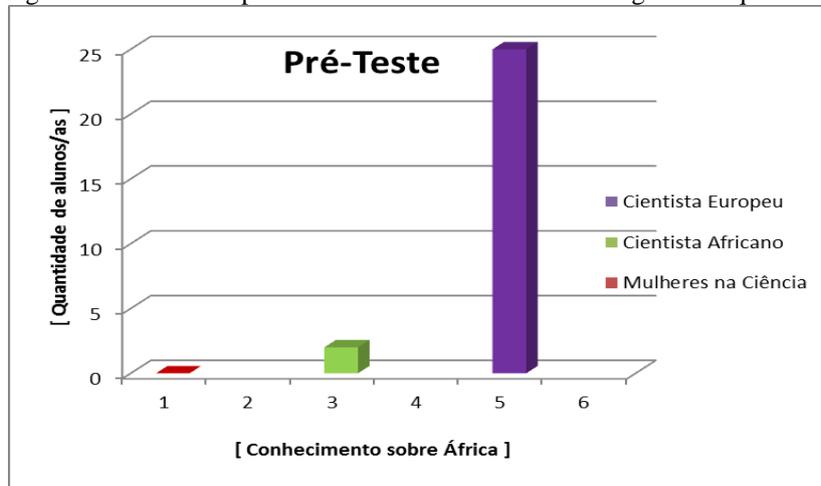

Fonte: (Os autores, 2017).

Figura 3: Pré-teste-físicos de diferentes continentes seguindo o quadro 1.

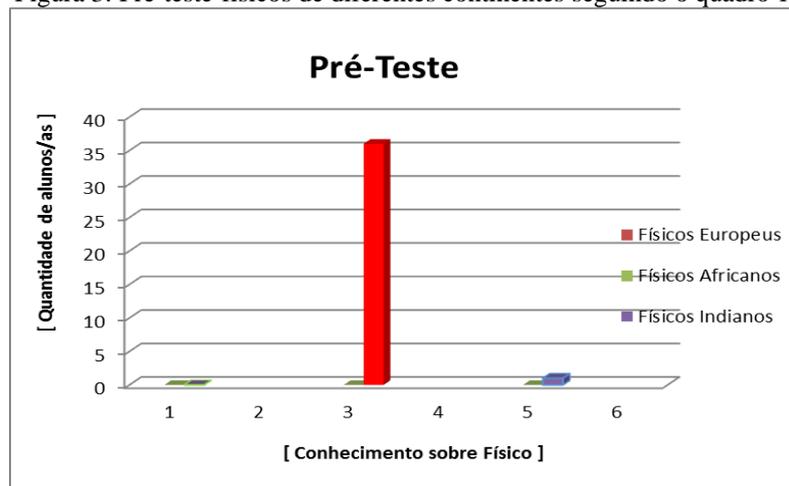

Fonte: (Os autores, 2017).

Em sequência à metodologia já descrita, após os seminários de aprofundamentos em temas relacionados à ciência e à cultura africana, foram feitas perguntas com as mesmas temáticas e, observou-se que houve um acúmulo e entendimento dos assuntos trabalhados em sala de sala. Alguns até ficaram espantados tamanha foi repercussão nos debates, tendo vista que as atividades estavam rompendo paradigmas. Desta forma, foi possível constatar uma crescente evolução - vide figuras 4, 5 e 6.

Figura 4: Pós-teste-conhecimentos sobre a África seguindo o quadro 1.

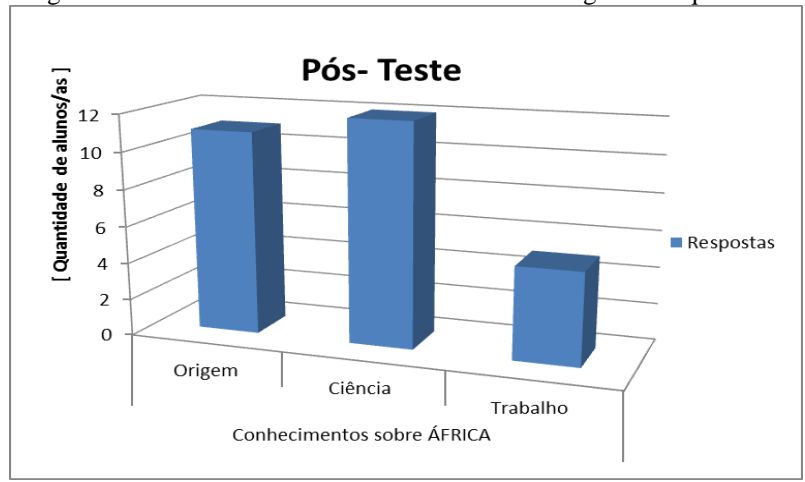

Fonte: (Os autores, 2017).

Figura 5: Pós-teste-respostas dos estudantes com o tema seguindo o quadro 1.

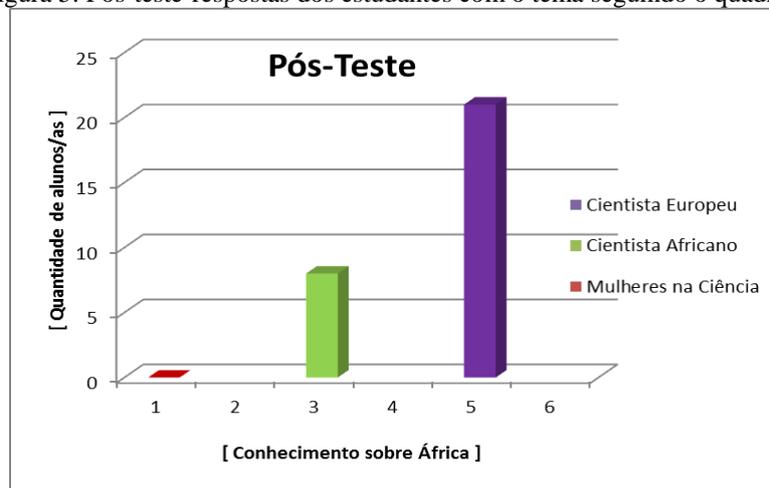

Fonte: (Os autores, 2017).

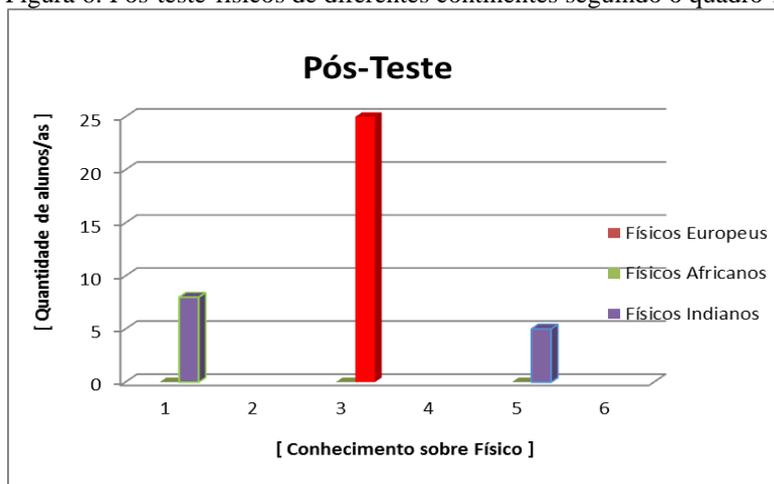

Figura 6: Pós-teste-físicos de diferentes continentes seguindo o quadro 1.

Fonte: (Os autores, 2017).

## 6. CONSIDERAÇÕES FINAIS

Algumas considerações práticas e pedagógicas precisam estabelecer compromisso em relação à promoção de uma educação étnico-racial. Nossos resultados (preliminares) revelam que os desafios são muitos, mas é preciso revalidar o fazer pedagógico inserido nos desafios das mudanças de valores, de lógicas e de representações sobre o outro, principalmente, aqueles que fazem parte dos grupos historicamente excluídos da sociedade.

Nessa perspectiva, nossa proposta cumpre seu papel formativo e se constitui numa estratégia para capacitar futuros professores de Física a não só produzirem materiais de ciências, tendo o cuidado de integrar os recursos de comunicação e informação, mas também materiais que atendam a uma demanda da legislação (lei 10.639/2003), oportunizando tratar, na sala de aula, temáticas étnico-raciais.

É urgente o fato de que nós, professores de Física, devemos nos posicionar diante dessa luta histórica e, portanto, em favor de práticas pedagógicas multiculturais no ambiente escolar, Precisamos assumir um compromisso político explícito frente às questões raciais que conduzam a uma educação antirracista.

O conhecimento científico não pode se limitar apenas às salas de aula, mas deve também contribuir significativamente na construção da identidade brasileira, permitindo a releitura de visões hegemônicas do mundo. Assim, podemos afirmar que essa investigação nos insere enquanto profissionais comprometidos com esta pauta e seus desdobramentos.

Vale ressaltar que houve relevante aprendizado junto à produção de conhecimento, no entanto, o projeto sinaliza dificuldades em evidenciar mulheres na ciência. Nesse sentido, a partir deste apontamento, esperamos despertar o interesse de continuidade de outros projetos pedagógicos que busquem investigar tal demanda, considerando a importância e necessidade de pensar estas questões pertinentes ao debate de raça e gênero.